\documentclass[epj,final]{svjour}
\smartqed  
\RequirePackage{graphicx}
\begin{document}

\title{Scaled variables and the quark-hadron duality}

\author{A.S.~Parvan\inst{1,}\inst{2}}

\institute{\inst{1}Bogoliubov Laboratory of Theoretical Physics, Joint Institute for Nuclear Research, Dubna, Russia \\ \inst{2}Department of Theoretical Physics, Horia Hulubei National Institute of Physics and Nuclear Engineering, Bucharest-Magurele, Romania}

\date{Received: date / Revised version: date}

\abstract{The thermodynamic quantities of the ideal gas of hadrons and the $(2+1)$--flavor lattice QCD scaled by the effective degeneracy factors of the corresponding models are compared. We have found that in terms of the scaled variables the quark-hadron duality of the lattice QCD and the hadron resonance gas (HRG) model disappears. However, we have unexpectedly revealed that the scaled variables lead to the quark-hadron duality of the lattice QCD and the quantum ideal gas of kaons and antikaons, namely, the ideal gas of those hadrons that contain all the three quarks $u,d,s$ and their antiquarks. Satisfactory agreement between the scaled results of the kaon ideal gas and the lattice QCD data is achieved at large values of the volume in the entire temperature range. In the ideal gas of kaons there is no any phase transition. Nevertheless, in our calculations the scaled thermodynamic quantities of the ideal gas and the lattice QCD follow the same qualitative behavior and are consistent with each other especially at high temperatures in the perturbative region.
} 
\titlerunning{Scaled variables and the quark-hadron duality}
\authorrunning{A.S.~Parvan}


\maketitle

\section{Introduction}
Perturbative calculations of low-energy hadronic observables from Quantum Chromodynamics (QCD), the theory of strong interactions, are not possible due to the fact that QCD does not contain a small dimensionless coupling constant. Thus, the lattice QCD (LQCD)~\cite{Miller2007}, which is one of the approaches of this theory, is used to proceed by defining QCD on a lattice of spacetime points, and then extrapolating observables to a continuum limit. In the LQCD calculations there is no any hadronic degree of freedom. This approach is formulated entirely in terms of quark and gluon fields. The main thermodynamic quantities, which are calculated in the framework of the LQCD, are pressure, energy density and entropy density. Their temperature and density dependence commonly summarized as the equation of state (EoS) of quarks and gluons~\cite{Bazavov2014,Borsanyi2014,Bazavov2017,Philipsen2013,Burger2015,Hegde2014,Caselle2018} provides the most basic description of the equilibrium thermodynamic properties of strong-interaction matter.

The fastest growth of reduced energy density $\varepsilon/T^{4}$ with temperature in LQCD is interpreted as a phase transition from the hadronic phase to the state of quark-gluon plasma (QGP)~\cite{Aoki2006}. Since LQCD theory does not contain hadronic degrees of freedom, the existence of the hadronic phase state is deduced indirectly from a comparison of the bulk thermodynamic quantities of LQCD with the results of other thermodynamic models, e.g. the hadron resonance gas (HRG) model~\cite{Ratti2011,Bazavov2017,Karsch2014,Andronic2012,Andronic2018,Vovchenko2015}. Connection of hadrons with LQCD from the comparison of the equations of state of the LQCD and HRG models at low temperatures is known as quark-hadron duality or hadron-parton duality~\cite{Andronic2018}. Namely, this coincidence of the equations of state is supposed to prove that LQCD predicts the existence of the hadron phase at low temperatures.

In the LQCD calculations and the HRG model, we have two different physical systems with a different number of types of particles or fields and different degrees of freedom. Thus, these two different physical systems are not comparable at the microscopic level. However, at the level of the macroscopic thermodynamic quantities (observables) there are two possibilities to compare them properly. First, this is a direct comparison of energy densities of the LQCD and HRG models as in refs.~\cite{Ratti2011,Bazavov2017,Karsch2014,Andronic2012,Andronic2018,Vovchenko2015}. Another possibility, which is the subject of the present paper, is to compare the scaled thermodynamic quantities, which are put to each other in correspondence. In the present paper, the scaled thermodynamic quantities of the model are defined as ratios of the corresponding thermodynamic quantities to the model effective degeneracy factor. We follow the standard definition of the effective degeneracy factor in the statistical mechanics given, for example, in refs.~\cite{Yagi2008,Brown,Turko}. That definition is compatible with the foundations of the statistical mechanics as it is independent of the thermodynamic variables of state.

The main aim of the present study is to analyze the thermal lattice QCD data in terms of quantum ideal gas of hadrons using the scaled variables of both these models: the lattice QCD and the hadron ideal gas.

In the present paper, we show that if we divide the caloric EoS of each model by its effective degeneracy factor, then the quark-hadron duality of the LQCD and HRG models disappears. However, the caloric EoS of the ideal gas of kaons divided by its degeneracy factor surprisingly qualitatively describes the EoS of LQCD in in the entire range of temperature $T$. It should be stressed that in the ideal gas of hadrons there is no phase transition.

The structure of the paper is as follows. In sect. 2, we discuss the ideal gas of hadrons. In sect. 3, we define the scaled variables and discuss the quark-hadron duality. The main conclusions are summarized in the final section.

\section{The ideal gas of hadrons}
For the ideal gas of bosons and fermions the density of the thermodynamic potential of the grand canonical ensemble in a finite volume $V$ can be written as
\begin{equation}\label{1}
  \omega = -\frac{1}{\beta V\eta} \sum\limits_{\vec{p},\sigma} \ln\left[1+\eta e^{-\beta(\varepsilon_{\vec{p}}-\mu)} \right],
\end{equation}
where $\beta=1/T$, $\eta=1$ for the Fermi-Dirac statistics and $\eta=-1$ for the Bose-Einstein statistics. Here, the hadron chemical potential $\mu=b\mu_{B}+q\mu_{Q}+s\mu_{S}$, where $\mu_{B},\mu_{Q},\mu_{S}$ and $b,q,s$ are the  baryonic, electric and strangeness chemical potentials of the system and the corresponding charges of hadron, respectively. The dispersion relation is
\begin{eqnarray}\label{2}
  \varepsilon_{\vec{p}} &=& \sqrt{\vec{p}^{2}+m^{2}}, \\ \label{3}
  p_{\alpha} &=& \frac{2\pi}{L} k_{\alpha}, \quad k_{\alpha}=0,\pm 1, \ldots, \quad \alpha=1,2,3,
\end{eqnarray}
where $L=V^{1/3}$ and $m$ is the mass of the hadron.

The mean occupation numbers for the ideal gas of hadrons in the grand canonical ensemble can be written as
\begin{equation}\label{4}
  \langle n_{\vec{p}\sigma}\rangle = \frac{1}{e^{\beta(\varepsilon_{\vec{p}}-\mu)}+\eta}.
\end{equation}
Then the density of hadrons and the energy density in a finite volume $V$ are
\begin{eqnarray}\label{5}
   \varrho &=& \frac{1}{V}\sum\limits_{\vec{p},\sigma}  \langle n_{\vec{p}\sigma}\rangle,  \\ \label{6}
  \varepsilon &=& \frac{1}{V} \sum\limits_{\vec{p},\sigma} \varepsilon_{\vec{p}}  \langle n_{\vec{p}\sigma}\rangle.
\end{eqnarray}
The pressure and the trace anomaly in a finite volume $V$ take the form
\begin{eqnarray}\label{7}
   p &=& \frac{1}{3V}\sum\limits_{\vec{p},\sigma} \frac{\vec{p}^{2}}{\varepsilon_{\vec{p}}} \langle n_{\vec{p}\sigma}\rangle,  \\ \label{8}
  \beta^{4}(\varepsilon-3p) &=& \frac{\beta^{4}}{V} \sum\limits_{\vec{p},\sigma} \frac{m^{2}}{\varepsilon_{\vec{p}}}  \langle n_{\vec{p}\sigma}\rangle.
\end{eqnarray}
The entropy density can be written as
\begin{equation}\label{9}
  s=-\beta (\omega - \varepsilon +\mu \varrho).
\end{equation}
For the ideal gas of fermions and bosons the density of the heat capacity in a finite volume $V$ is
\begin{equation}\label{10}
  c_{V\mu} = \frac{\beta^{2}}{V} \sum\limits_{\vec{p},\sigma} (\varepsilon_{\vec{p}}-\mu)^{2}  \langle n_{\vec{p}\sigma}\rangle (1-\eta \langle n_{\vec{p}\sigma}\rangle).
\end{equation}

\begin{figure*}
\includegraphics[width=0.98\textwidth]{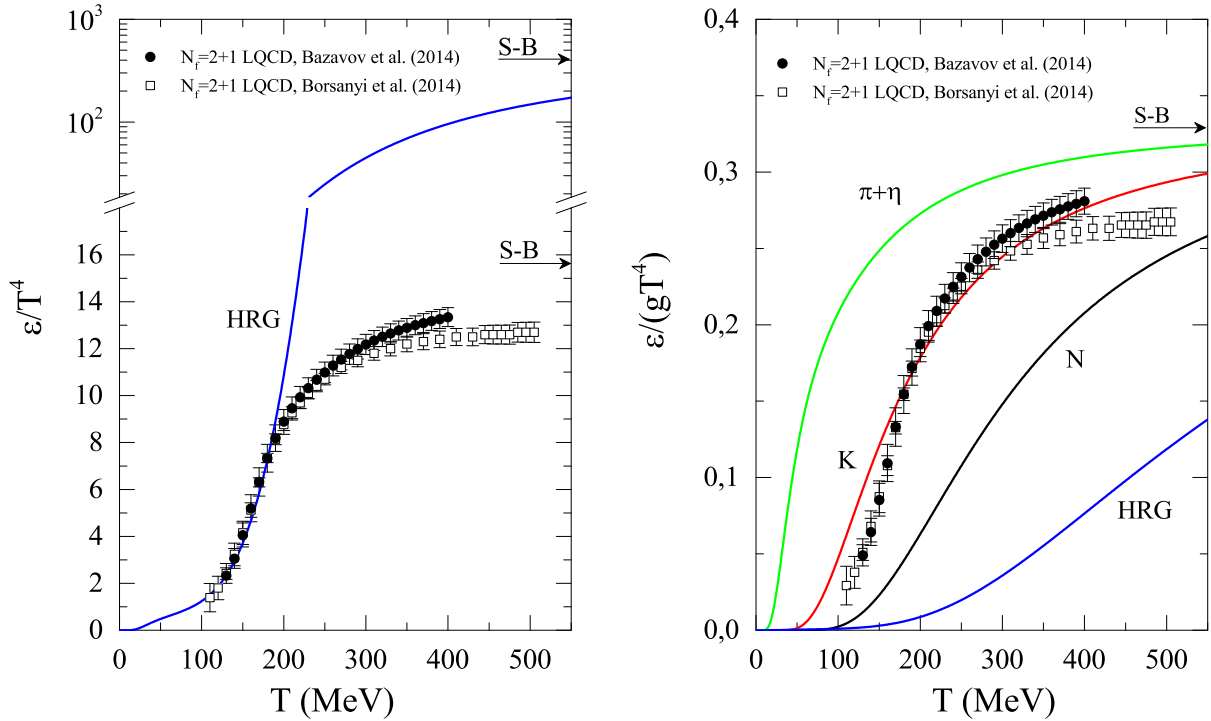}
\caption{(Color online) Left panel: Comparison of the prediction of the HRG model (ideal gas of hadrons) with the lattice QCD equation of state with $N_{f}=2+1$ flavors obtained by two different collaborations~\cite{Bazavov2014,Borsanyi2014}. The calculations for the HRG model were carried out in the volume $V=10^{3}$ fm$^{3}$ at $\mu=0$. The arrows represent the Stefan-Boltzmann limit for the energy density of the ideal gas of quarks and gluons, $95\pi^{2}/60$, and the ideal gas of hadrons. Right panel: the scaled energy density as a function of temperature $T$ for the ideal gas of hadrons and for the lattice QCD with $N_{f}=2+1$ flavors~\cite{Bazavov2014,Borsanyi2014}. The lines $\pi+\eta$, $K$, $N$ and HRG correspond to the calculations for the ideal gas of light unflavored mesons $\pi^{\pm},\pi^0,\eta$, kaons $K^{\pm},K^0,\bar{K}^0$, nucleons $p,n,\bar{p},\bar{n}$ and for the HRG model in the volume $V=10^{3}$ fm$^{3}$ at $\mu=0$. The arrow represents the Stefan-Boltzmann limit for the scaled energy density, $\pi^{2}/30$. Symbols represent the lattice QCD data. }  \label{fig1}
\end{figure*}

\section{Scaling Duality}
The left panel of Fig.~\ref{fig1} shows the dependence of the energy density $\varepsilon$ on the temperature $T$ for the hadron resonance gas model and the lattice QCD with $N_{f}=2+1$ flavors obtained by two different collaborations~\cite{Bazavov2014,Borsanyi2014}. The equation of state for the HRG model is calculated in the volume $V=10^{3}$ fm$^{3}$ at the chemical potential $\mu=0$. The particle masses are taken from the Particle Data Group~\cite{pdg}, including all known states up to the resonance masses of $f_{0}(1500)$ for the light unflavored mesons, $K_{2}^{*}(1430)$ for the strange mesons, $N(2600)$ for $N$ baryons, $\Delta(2420)$ for the delta baryons, $\Lambda(2350)$ for the lambda baryons, $\Sigma(2250)$ for the sigma baryons, $\Xi(2030)$ for the xi baryons and $\Omega(2250)^{-}$ for the omega baryons. We have obtained the same results as in~\cite{Bazavov2014} (see the left panel of Fig.~\ref{fig1}). The HRG model describes very well the lattice QCD calculations for the caloric equation of state (the dependence of the energy of the system on temperature) with $N_{f}=2+1$ flavors in the region of low temperatures. It is considered that this coincidence reflects the existence of the quark-hadron duality~\cite{Andronic2018}. Let us redraw the same dependence of the energy density $\varepsilon$ on temperature $T$ on a larger scale (see on the top of the left panel of Fig.~\ref{fig1}). From this picture it is clearly seen that the caloric equation of state of the HRG model has similar qualitative behavior as the caloric equation of state of the lattice QCD but on different scales. At high temperatures the energy density of the lattice QCD tends to its Stefan-Boltzmann limit and the energy density of the ideal hadron gas tends to the Stefan-Boltzmann limit of the HRG model. The difference in the Stefan-Boltzmann limit of the HRG model and the lattice QCD lies in the different number of species of particles or fields and their different degrees of freedom in these two models. Therefore, let us scale the energy densities of the HRG model and the lattice QCD by their own effective degeneracy factors.

The effective degeneracy factor of the quark and gluon fields can be written as~\cite{Yagi2008,Brown,Turko}
\begin{eqnarray}\label{1a}
  g_{QCD} &=& g_{g}+\frac{7}{8}g_{q}, \\ \label{2a}
  g_{g} &=& 2_{spin} \times (N_{c}^{2}-1), \\ \label{3a}
  g_{q} &=&  2_{spin} \times  2_{q\bar{q}} \times N_{c} \times N_{f}.
\end{eqnarray}
The factor $7/8$ in eq.~(\ref{2a}) appears from the difference of the Bose-Einstein and Fermi-Dirac statistics. The effective degeneracy factor of the hadron gas can be defined as
\begin{equation}\label{4a}
  g_{H}=\sum\limits_{M} g_{M} + \frac{7}{8} \sum\limits_{B} g_{B},
\end{equation}
where $g_{i}=(2J_{i}+1)(2I_{i}+1)$ is the spin-isospin degeneracy factor of the (anti)mesons $(i=M)$ and (anti) baryons $(i=B)$. In the present study, we use the simplest definition for the effective degeneracy factors of both these physical systems in order to preserve the essence of the results of the lattice QCD.

\begin{figure}
\includegraphics[width=0.45\textwidth]{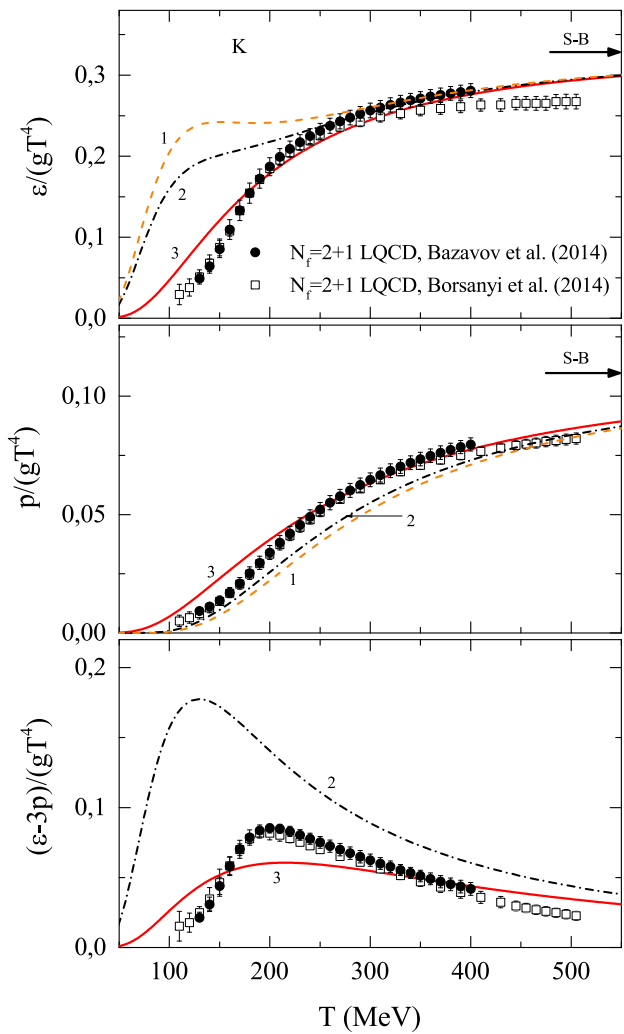}
\caption{(Color online) The scaled energy density, the scaled pressure and the scaled trace anomaly as functions of temperature $T$ for the ideal gas of kaons $(K^{\pm},K^0,\bar{K}^0)$ and for the lattice QCD with $N_{f}=2+1$ flavors~\cite{Bazavov2014,Borsanyi2014}. The lines $1,2$ and $3$ are calculations for the ideal gas in the volume $V=1.1^{3},1.2^3$ and $10^{3}$ fm$^{3}$, respectively, at $\mu=0$. The arrows represent the Stefan-Boltzmann limit for the scaled energy density, $\pi^{2}/30$, and the scaled pressure, $\pi^{2}/90$. Symbols represent the lattice QCD data. }  \label{fig3}
\end{figure}

Let us divide the energy densities of the HRG model and the lattice QCD by the effective degeneracy factors $g_H$ (\ref{4a}) and $g_{QCD}$ (\ref{1a}), respectively, and perform the same calculations for the ideal gases of some other individual species of hadrons. The right panel of Fig.~\ref{fig1} represents the dependence of the scaled energy density $\varepsilon/g$ on the temperature $T$ for the HRG model $(g_{HRG}=1251)$, the ideal gas of nucleons and antinucleons $p,n,\bar{p},\bar{n}$ $(g_N=7)$, kaons and antikaons $K^{\pm},K^0,\bar{K}^0$ $(g_K=4)$, light unflavored mesons $\pi^{\pm},\pi^0,\eta$ $(g_{\pi+\eta}=4)$ and for the lattice QCD with $N_{f}=2+1$ flavors $(g_{QCD}=47.5)$ obtained by two different collaborations~\cite{Bazavov2014,Borsanyi2014}. The calculations for the ideal gas of hadrons were performed in the volume $V=10^{3}$ fm$^{3}$ at the chemical potential $\mu=0$. Now the scaled energy density of the ideal gas of hadrons and the lattice QCD have the same Stefan-Boltzmann limit, $\pi^{2}/30$. It is clearly seen that the scaled energy density of the HRG model differs essentially from the scaled energy density of the lattice QCD in the entire range of $T$. It increases very slowly with $T$. The scaled energy density of the ideal gas of low lying nucleons and antinucleons underestimates the lattice QCD data, but the scaled energy density of the ideal gas of low lying light unflavored mesons overestimates the lattice QCD results. However, we have found that only the scaled energy density of the ideal gas of low lying kaons and antikaons describes very well the data of the lattice QCD with $N_{f}=2+1$ flavors. The curve of the ideal gas of $K^{\pm}$, $K^{0}$ and $\bar{K}^{0}$ mesons has similar qualitative behaviour as the lattice QCD caloric equation of state and it approaches the lattice QCD data at temperatures higher than $170$ MeV in the region of deconfinement of quarks and gluons.

Why does only the ideal gas of kaons and antikaons describe very well the lattice QCD scaled data with $N_{f}=2+1$ flavors? This can be explained by the fact that only kaons and antikaons contain all three quarks $u,d,s$ and antiquarks $\bar{u},\bar{d},\bar{s}$ which were included in the lattice calculations with $N_{f}=2+1$ flavors~\cite{Bazavov2014,Borsanyi2014}. The pions, $\eta$ meson and nucleons contain $u$ and $d$ quarks but they do not contain $s$ and $\bar{s}$ quarks. Moreover, it should be stressed that only kaons and antikaons with the smallest masses can describe the lattice QCD data. The inclusion of the exited kaons like $K^{*}$ leads to large deviations of the scaled energy density of the ideal gas from the lattice QCD result.

Let us study the finite volume corrections to the thermodynamic quantities of the ideal gas of kaons and antikaons in a finite volume. Figure~\ref{fig3} represents the behavior of the scaled energy density, the scaled pressure and the scaled trace anomaly as functions of temperature $T$ for the ideal gas of kaons and antikaons in the different volumes $V$ at $\mu=0$. The lines $1,2$ and $3$ correspond to the calculations for the ideal gas in the volume $V=1.1^{3},1.2^3$ and $10^{3}$ fm$^{3}$, respectively. The symbols represent the lattice QCD data~\cite{Bazavov2014,Borsanyi2014}. The arrows represent the Stefan-Boltzmann limit for the scaled energy density, $\pi^{2}/30$, and the scaled pressure, $\pi^{2}/90$. For the small values of volume $V=1.2^{3}$ fm$^3$ and at high temperatures $T>250$ MeV, the scaled energy density of the ideal gas of kaons and antikaons describes very well the lattice QCD data obtained by the HotQCD Collaboration~\cite{Bazavov2014}. However, at small values of temperature and small values of volume the scaled energy density of the ideal gas deviates essentially from the lattice QCD results. For large values of the volume $V=10^{3}$ fm$^{3}$, the scaled energy density of the ideal gas of kaons and antikaons becomes closer to the lattice QCD equation of state in the region of small temperatures $T$ and describes it very well in the region of high temperatures at $T>170$ MeV. The behavior of the scaled energy density of the ideal gas of kaons and antikaons in a large volume is the same as the behavior of the scaled energy density of the lattice QCD.

For small values of the volume $V=1.2^{3}$ fm$^3$ and at $T<150$ and $T>400$ MeV, the scaled pressure of the ideal gas of kaons and antikaons is very close to the lattice QCD data. See the middle panel of Fig.~\ref{fig3}. For large values of the volume $V=10^{3}$ fm$^{3}$, the scaled pressure of the ideal gas of kaons and antikaons describes very well the lattice QCD scaled pressure in the region of high temperatures at $T>200$ MeV and does not differ essentially from it in the region of small temperatures $T$. The behavior of the scaled pressure of the ideal gas of kaons and antikaons in a large volume is the same as the behavior of the scaled pressure of the lattice QCD.

The behavior of the scaled trace anomaly of the ideal gas of kaons and antikaons in both large and small volumes is similar to the behavior of the scaled trace anomaly of the lattice QCD. See the lower panel of Fig.~\ref{fig3}. However, the curves of the scaled trace anomaly of the ideal gas differ quantitatively from the results of the lattice QCD. The scaled trace anomaly of the ideal gas of kaons and antikaons decreases with the volume $V$.

\begin{figure}[pt]
\includegraphics[width=0.45\textwidth]{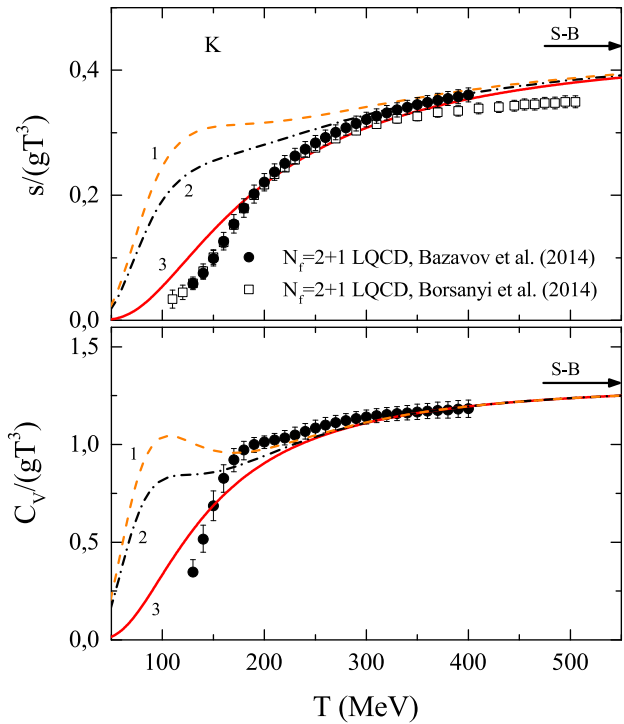}
\caption{(Color online) The scaled entropy density and the scaled heat capacity as functions of temperature $T$ for the ideal gas of kaons $(K^{\pm},K^0,\bar{K}^0)$ and for the lattice QCD with $N_{f}=2+1$ flavors~\cite{Bazavov2014,Borsanyi2014}. The lines $1,2$ and $3$ are calculations for the ideal gas in the volume $V=1.1^{3},1.2^3$ and $10^{3}$ fm$^{3}$, respectively, at $\mu=0$. The arrows represent the Stefan-Boltzmann limit for the scaled entropy density, $2\pi^{2}/45$, and the scaled heat capacity, $2\pi^{2}/15$. Symbols represent the lattice QCD data. }  \label{fig4}
\end{figure}

Figure~\ref{fig4} shows the behavior of the scaled entropy density and the scaled heat capacity as functions of temperature $T$ for the ideal gas of kaons and antikaons in the different volumes $V$ at $\mu=0$. The lines $1,2$ and $3$ correspond to the calculations for the ideal gas in the volume $V=1.1^{3},1.2^3$ and $10^{3}$ fm$^{3}$, respectively. The symbols denote the lattice QCD data~\cite{Bazavov2014,Borsanyi2014}. The arrows represent the Stefan-Boltzmann limit for the scaled entropy density, $2\pi^{2}/45$, and the scaled heat capacity, $2\pi^{2}/15$. For small values of the volume $V=1.2^{3}$ fm$^3$ and at $T>250$ MeV, the scaled entropy density of the ideal gas of kaons and antikaons describes very well the lattice QCD scaled results obtained by the HotQCD Collaboration~\cite{Bazavov2014}. However, at small values of temperature the scaled entropy density of the ideal gas deviates substantially from the lattice QCD data. For large values of the volume $V=10^{3}$ fm$^{3}$, the scaled entropy density of the ideal gas of kaons and antikaons becomes closer to the lattice QCD scaled entropy density in the region of small temperatures $T$ and describes it very well in the region of high temperatures at $T>170$ MeV. The behavior of the scaled entropy density of the ideal gas of kaons and antikaons in a large volume is the same as the behavior of the scaled entropy density of the lattice QCD and they are very close between themselves. The high temperature plateau in the scaled heat capacity of the lattice QCD at $T>170$ MeV is well described by the ideal gas of kaons and antikaons in the volume $V=1.1^{3}$ fm$^3$. See the lower panel of Fig.~\ref{fig4}. For large values of the volume $V=10^{3}$ fm$^{3}$, the scaled heat capacity of the ideal gas of kaons and antikaons is compatible with the lattice QCD results in the region of high temperatures at $T>250$ MeV. In general, the behavior of the scaled heat capacity of the ideal gas of kaons and antikaons in a large volume is similar to  the behavior of the scaled heat capacity of the lattice QCD.

It should be stressed that the continuum limit of the thermodynamic quantities obtained by S.~Bors\'{a}nyi et al.~\cite{Borsanyi2014} at high temperatures underestimates the continuum limit results of the HotQCD Collaboration~\cite{Bazavov2014} and of the ideal gas of kaons. See, for instance, Figs.~\ref{fig3} and~\ref{fig4}.

\section{Conclusions}
In the present paper, the comparison of the equations of state of the ideal gas of hadrons and the lattice QCD with $N_{f}=2+1$ flavors has been performed. We reproduced the results of the HRG model on the dependence of the energy density on the temperature and confirmed the quark-hadron duality at low temperatures in the region of the confinement of quarks and gluons. However, after considering these results on a larger scale of temperatures and energies, we revealed that the caloric equations of state of the HRG model and the lattice QCD model have the same qualitative behavior and differ only in scale and the value of the Stefan-Boltzmann limit. Thus, these quantities are not commensurable since they correspond to different degrees of freedom. Therefore, in order to properly compare them, we bring them to the same scale reducing the degree of freedom by dividing the energy density to the effective degeneracy factor of the corresponding system. After applying this procedure, we obtained that the scaled energy density of the HRG model significantly deviates from the scaled energy density of the lattice QCD model in the entire range of temperature $T$. Thus, in terms of the scaled variables the quark-hadron duality of the HRG model disappears. However, we revealed that the scaled energy density of the lattice QCD with $N_{f}=2+1$ flavors can be well described only by an ideal gas of kaons and antikaons, namely, by the ideal gas of those hadrons that contain all three quarks $u,d,s$ and their antiquarks. The scaled caloric curve of the ideal gas of kaons and antikaons is similar to the lattice QCD caloric equation of state in the whole region of temperature; however, the best description of the lattice QCD data is achieved at high temperatures in the region of deconfinement of quarks and gluons. The ideal gas of nucleons and antinucleons and the ideal gas of low lying light unflavored mesons do not describe the scaled results of the lattice QCD with $N_{f}=2+1$ flavors. This is the result of the fact that these hadrons do not contain $s$ and $\bar{s}$ quarks. We have found also that only kaons and antikaons with the smallest masses can describe the lattice QCD data.

The finite volume corrections to the thermodynamic quantities of the kaon ideal gas were studied. We have found that the scaled thermodynamic quantities of the lattice QCD can be well described by the ideal gas of kaons only at large values of volume $V$. At small values of volume and temperature the results of the ideal gas differ substantially from the lattice QCD data. The scaled energy density, the scaled pressure and the scaled entropy of the $(2+1)$ - flavor lattice QCD are well described by the ideal gas of kaons and antikaons. However, the scaled trace anomaly of the ideal gas has a similar trend as the scaled trace anomaly of the lattice QCD but differs from it quantitatively. The scaled heat capacity of the ideal gas also has a similar trend as the scaled heat capacity of the lattice QCD and coincides with it at high temperatures only. In general, the scaled thermodynamic quantities of the ideal gas of kaons and antikaons are consistent with the scaled results of the lattice QCD with $N_{f}=2+1$ flavors and show a similar trend. However, it should be stressed that in the ideal gas of kaons and antikaons there are no any phase transitions.

{\bf Acknowledgments:} This work was supported in part by the joint research project of JINR and IFIN-HH. I am indebted to T.~Bhattacharyya, J.~Cleymans, S.~Mogliacci, A.S.~Sorin and O.V.~Teryaev for fruitful discussions. I am also grateful to M.~Ilgenfritz for the discussions related to the lattice QCD.

\end{document}